\documentclass[10pt]{article}
\usepackage{amsmath,amssymb}
\usepackage{graphicx}
 \allowdisplaybreaks
 \newcommand{\be}{\begin{equation}}
 \newcommand{\ee}{\end{equation}}
 \newcommand{\bea}{\begin{eqnarray}}
 \newcommand{\eea}{\end{eqnarray}}
 
 \newcommand{\nn}{\nonumber}
 \newcommand{\td}{\tilde}
 \newcommand{\wtd}{\widetilde}
 \newcommand{\pd}{\partial}
 \newcommand{\one}{{\bf 1}}

 \newcommand{\bk}{{\bf k}}

 \newcommand{\bw}{{\bf w}}
 
 \newcommand{\bJ}{{\bf J}}
 \newcommand{\bL}{{\bf L}}
 
 \newcommand{\bQ}{{\bf Q}}

 \newcommand{\cA}{{\cal A}}

 \newcommand{\cL}{{\cal L}}
 \newcommand{\cM}{{\cal M}}
 
 \newcommand{\cO}{{\cal O}}

 \newcommand{\lie}{\pounds}
 
 \newcommand{\dtp}{\dot{p}}
 \newcommand{\dtq}{\dot{q}}

\long\def\symbolfootnote[#1]#2{\begingroup%
\def\thefootnote{\fnsymbol{footnote}}\footnote[#1]{#2}\endgroup}

\newcommand{\aei}{\it Max Planck Institute for Gravitational Physics
(Albert Einstein Institute)\\ Am M\"uhlenberg 1, 14476 Golm,
Germany}

\newcommand{\auth}{Jianwei Mei}

\begin{document}
\thispagestyle{empty}
\begin{flushright}
\hfill{AEI-2012-016}
\end{flushright}
\begin{center}

~\vspace{20pt}

{\Large\bf On the General Kerr/CFT Correspondence in Arbitrary
Dimensions}

\vspace{25pt}

\auth \symbolfootnote[1]{Email:~\sf jwmei@aei.mpg.de}



\vspace{10pt}{\aei}

\vspace{2cm}

\underline{ABSTRACT}

\end{center}

We study conformal symmetries on the horizon of a general
stationary and axisymmetric black hole. We find that there exist
physically reasonable boundary conditions that uniquely determine
a set of symmetry generators, which form one copy of the Virasoro
algebra. For extremal black holes, Cardy's formula reproduces
exactly the Bekenstein-Hawking entropy.

 \newpage

\tableofcontents

\section{Introduction}\label{sec.intro}

There is an intriguing possibility that quantum gravity on the
horizon of a black hole could be dual to some 2D conformal filed
theory (CFT) at finite temperatures. A key evidence is that,
following \cite{brown.henneaux86}, one can find appropriate
boundary conditions which allow for asymptotic conformal
symmetries on the horizon
\cite{strominger97,carlip98,carlip99,ghss08}. In the case of
extremal black holes, it is possible to study the relevant
conformal symmetries in a systematic fashion \cite{mei10}, based
on the idea of the Kerr/CFT correspondence \cite{ghss08}. For
non-extremal black holes, it appears more difficult to study (or
identify) the relevant conformal symmetries directly. In
\cite{cms10} it has been proposed to use a probing scalar field to
extract useful information about the possible hidden conformal
symmetries in the Kerr background. The method has been further
developed and applied to several other cases (see, e.g.
\cite{cvetic.larsen11a} and references therein).

Such activity has also generated renewed interest in the earlier
effort \cite{carlip98,carlip99,park01} that seeks to study the
conformal symmetries on the horizons directly. In particular, it
has been shown in \cite{carlip11,chen.zhang11} that it is possible
to use the ``stretched horizon" method to re-derive some results
known in the usual literature of Kerr/CFT correspondence
\cite{ghss08}. What's more, Carlip \cite{carlip11} has shown that
in the case when the boundary conditions are the same for both the
extremal and non-extremal black holes, there is a Virasoro algebra
which reproduces (via the Cardy formula) the full entropy for
extremal black holes but only half the entropy for non-extremal
ones. And the same paper also shows that an alternative way of
stretching the horizon can enable one to obtain the entropy fully.
In view of such ambiguities, it will be helpful to study boundary
symmetries on the horizon without using an intermediate stretched
horizon \cite{carlip11}. Our main purpose of this work is to
present such a construction.

We find physically reasonable boundary conditions that uniquely
determine a set of symmetry generators, which form one copy of the
Virasoro algebra (one for each of the azimuthal angles). Cardy's
formula can then be used to calculate the black hole entropy. Our
boundary conditions are directly imposed on the (inverse) metric
elements on the horizon, in much the same spirit as Brown and
Henneaux \cite{brown.henneaux86}. The construction is general and
is valid for arbitrary stationary and axisymmetric black holes in
arbitrary dimensions. For practical reasons, some of the
calculation is only explicitly done for Einstein gravity plus a
(possibly zero) cosmological constant. But with enough effort, a
generalization to more complicated theories should still be
possible.

The main result of the paper is the following:

In section \ref{sec.bh}, we briefly recall some general features
of stationary and axisymmetric black holes, setting the stage for
our discussion. In section \ref{sec.Vira}, we explain our boundary
conditions, solve for the boundary symmetry generators, showing
that they constitute a copy of the Virasoro algebra. In section
\ref{sec.non.extremal}, we calculate the central charge and the
Frolov-Thorne temperature for non-extremal black holes. We find
that Cardy's formula gives exactly half the Bekenstein-Hawking
entropy. In section \ref{sec.extremal}, we do the same for
extremal black holes. Here we find that Cardy's formula reproduces
the Bekenstein-Hawking entropy fully. We end with a short summary
in section \ref{sec.summary}.

Both the definition of black hole charges and the calculation of
central charges are done by using the covariant phase space
method, for which we collect some basic formulae in the appendix
\ref{sec.app}.

\section{Stationary and axisymmetric black holes}\label{sec.bh}

Stationary and axisymmetric black holes constitute the most
important class of exact solutions in various gravitational
theories. They are the objects that we want to focus on in this
paper. For the convenience of later discussions and also to fix
our notations, we briefly recall some general features of these
black holes. Most result has already appeared in \cite{mei11}, but
here we shall explain some of the points in more detail.

In general, a stationary and axisymmetric black hole is
characterized by the presence of a time-like Killing vector
$\pd_t$ and one (or several) space-like Killing vector(s)
$\pd_\phi$, where $\phi$ is periodically identified. Although a
general proof is not known, existing examples suggest that all the
stationary and axisymmetric black holes share the following form
of the metric,\footnote{A careful check of the general metric
against many existing examples can be found in \cite{mei10}. (The
metrics in \cite{mei10} look slightly different, but it is easy to
put them into the form of (\ref{metric.general}).) As a convention
for our sub/superscripts, the beginning Latin letters
$(a,b,\cdots)$ are only used for the azimuthal angles (e.g.,
$\phi^a$), the middle Latin letters $(i,j,\cdots)$ are only used
for the longitudinal angles (e.g., $\theta^i$), and the Greek
letters $(\mu,\nu,\cdots)$ are used for all the coordinates,
$\mu,\nu,\cdots\in\{r,t,a,i\}$. What's more, it is often
convenient to treat the time on the same footing as the azimuthal
angles. So we define $w^t=\Omega^t\equiv1$ and we use capital
letter indices ($A,B,\cdots$) to go over both the azimuthal angles
and the time, i.e. $A,B,\cdots\in\{t,a\}$. (In the brackets
$\{r,t,a,i\}$ and $\{t,a\}$, we use ``$a$" to represent all
indices for the azimuthal angles and ``$i$" to represent all
indices for the longitudinal angles.)}
\be ds^2=f\Big[-\frac{\Delta}{v^2}dt^2+\frac{dr^2}\Delta\Big]
+q_{ij}d\theta^i d\theta^j+g_{ab}(d\phi^a-w^adt)(d\phi^b-w^bdt)
\,,\label{metric.general}\ee
where the coordinates can be identified as $t$ the asymptotic
time, $r$ the radial coordinate, $\theta^i$ the longitudinal
angles and $\phi^a$ the azimuthal angles. All the functions in
(\ref{metric.general}) depend on $r$ and $\theta^i$, except for
$\Delta$ which is only a function of $r$. The inverse of
(\ref{metric.general}) is
\be (\pd_S)^2=\frac{\Delta}{f}\pd_r^2 +q^{ij} \pd_i \pd_j
+g^{ab}\pd_a\pd_b -\frac{v^2}{f \Delta} (\pd_t+w^a\pd_a)
(\pd_t+w^b \pd_b)\,, \label{metric.inverse}\ee
where $\pd_i\equiv\pd_{\theta^i}$ and $\pd_a\equiv\pd_{\phi^a}$.
Using $\td{g}_{\mu\nu}$ to denote elements of the full metric, we
find
\bea \td{g}_{rr}=\frac{f}\Delta\,,\quad \td{g}_{ij}=q_{ij}\,,
\quad \td{g}_{ab}=g_{ab}\,, \quad \td{g}_{at} =-w_a\,,\quad
\td{g}_{tt}=-N^2+w^2\,,\nn\\
\td{g}^{rr}=\frac\Delta{f}\,,\quad \td{g}^{ij}=q^{ij}\,,\quad
\td{g}^{ab}=g^{ab}-\frac{w^aw^b}{N^2}\,,\quad \td{g}^{at}=-
\frac{w^a}{N^2}\,,\quad \td{g}^{tt}=-\frac1{N^2}\,,
\label{elements.metric}\eea
where $N^2=f\Delta/v^2$, $w_a=g_{ab}w^b$ and $w^2=w_aw^a$. The
determinant of the full metric is $\td{g}=-q g f^2/v^2$, where $q$
is the determinant of $q_{ij}$ and $g$ is the determinant of
$g_{ab}$.

The (outer) black hole horizon $r_0$ is located at the (largest)
root of $\Delta(r_0)=0$. Near the black hole horizon, $f, v^2,
(q_{ij})$ and $(g_{ab})$ are all positive definite. The fact that
black holes are intrinsically regular on the horizon puts extra
constraints on the functions,
\bea v(r,\theta^i)&=&v_0(r)+v_1(r,\theta^i)\Delta+\cO(\Delta^2)\,,\nn\\
w^a(r,\theta^i)&=&w_0^a(r)+w^a_1(r,\theta^i)\Delta+\cO(\Delta^2)\,,
\label{wa.expansion}\eea
which means that any dependence of $v$ and $w^a$ on $\theta^i$ can
only begin at the order $\Delta$. With these conditions, it is
then possible to completely remove the divergence at $\Delta
\rightarrow0$ from the metric,
\bea -\frac{\Delta}{v^2}dt^2+\frac{dr^2}\Delta=\Big(\frac{
\Delta}{v_0^2}-\frac{\Delta}{v^2}\Big)dt^2 -\frac{\Delta}{
v_0^2}du_+ du_-\,,\nn\\
d\phi^a-w^adt=d\phi_\pm^a -w^a du_\pm\pm (w^a-w_0^a)\frac{v_0}
\Delta dr\,,\eea
where
\be du_\pm =dt\pm \frac{v_0}\Delta dr\,,\quad d\phi_\pm^a =d\phi^a
\pm w_0^a \frac{v_0}\Delta dr\,. \label{upm.and.phipm}\ee
The constraints that $w_0^a$ and $v_0$ depend only on $r$ comes
from the required integrability of (\ref{upm.and.phipm}). In the
presence of matter fields, similar constraints should also apply.
For example, if there is a $U(1)$ gauge field, then it must be of
the form
\be A=A_a(r,\theta^i)(d\phi^a-w^adt)+\Big[A_t(r)
+\cO(\Delta)\Big]dt\,.\ee

Finally, one can choose the coordinate system to be non-rotating
at spatial infinity, which means
\be w^a(r,\theta^i)\quad\longrightarrow\quad 0\quad{\rm as}\quad
r\rightarrow+\infty\,.\label{wa.infty}\ee
In this case, $\Omega^a=w_0^a(r_0)$ is the angular velocity of the
horizon along $\phi^a$. Using the null Killing vector on the
horizon $\pd_t+\Omega^a\pd_a$, one can find that the black hole
temperature is
\be T=\frac\kappa{2\pi}=\frac{\Delta'}{4\pi v} \Big|_{r=r_0}
=\frac{\Delta'(r_0)}{4\pi v_0(r_0)}\,,\label{temperature}\ee
where $\kappa$ is the surface gravity on the horizon. For extremal
black holes, $\Delta'(r_0)=0$ and so the temperature vanishes.

Charges of the black hole can be calculated by using
(\ref{app.def.Hxi2}) in the appendix,
\bea\delta E&=&\int_{horizon}\delta\bQ_{(\pd_t)}
-i_{(\pd_t)}{\bf\Theta}_{\delta}\,,\label{def.E}\\
\delta J_a&=&-\int_{horizon}\delta\bQ_{(\pd_a)}
-i_{(\pd_a)}{\bf\Theta}_{\delta} =-\int_{horizon}
\delta\bQ_{(\pd_a)}\,. \label{def.Ja}\eea
In these definitions, the charges will not be well defined unless
the corresponding defining equations are $\delta$-integrable. For
$J_a$, the $\delta$-integrability of (\ref{def.Ja}) is obvious and
one has
\be J_a=-\int_{horizon}\bQ_{(\pd_a)}\,.\ee
For $E$, one can use (\ref{metric.general}) to explicitly check
\cite{mei11} that in the context of Einstein gravity plus a
(possibly zero) cosmological constant,\footnote{The operator
$\bar\delta$ is defined to perturb only the free parameters (such
as mass and angular momenta) in a given solution. This is what's
usually needed to test the first law of thermodynamics for a given
black hole solution. For all other types of perturbations, one can
consult \cite{wald93,iyer.wald94}.}
\be\bar\delta E=T\bar\delta S +\Omega^a\bar\delta J_a\,,\quad
S=\frac{\cA_{rea}}4\,, \label{first.law}\ee
where $S$ is the Bekenstein-Hawking entropy and $\cA_{rea}\equiv
\int_{horizon}(d^{D-2}x)_{tr}2\sqrt{q g}\;$ is the area of the
horizon. It is obvious that the $\bar\delta$-integrability of
(\ref{def.E}) is intimately related to the presence of the first
law of thermodynamics (\ref{first.law}) for the black holes. In
fact, the authors of \cite{gibbons.perry.pope04} have noticed
integrating the first law of thermodynamics as a practical method
for calculating the mass of black holes, especially for ones that
do not have an easier alternative.

\section{Boundary conditions and the Virasoro algebra}\label{sec.Vira}

As is obvious from the last section, all thermodynamical
quantities of a black hole can be calculated purely by using data
from the neighborhood of the horizon. From this perspective, two
black holes are {\it intrinsically} the same if they approach each
other fast enough as one takes the limit to the horizon, while the
exterior of the black holes may be rather different due to matter
fields living outside the horizon. This is our most important
reason for choosing to impose boundary conditions on the horizon.
A loosely related and interesting idea can be found in
\cite{cvetic.larsen11a, cvetic.larsen11b, cvetic.gibbons12}.

The fluctuations over a given background $\td{g}_{\mu\nu}$ should
satisfy the linearized equations of motion, which we denote as
\be\td{E}_{\mu\nu}=0\,,\quad\Longrightarrow\quad
\delta\td{E}_{\mu\nu}=0\,.\ee
A particular class of solutions are generated by a Lie derivative,
\be\delta \td{g}_{\mu\nu} =\lie_\xi \td{g}_{\mu \nu}\,,\quad
\Longrightarrow \quad \delta\td{E}_{\mu\nu}=\lie_\xi
\td{E}_{\mu\nu}=0\,.\ee
This may not be much a surprise, because $\delta \td{g}_{\mu\nu}
=\lie_\xi \td{g}_{\mu \nu}$ (when paired with $\delta
x^\mu=-\xi^\mu$) is nothing but the general diffeomorphism.
However, the new metric $\td{g}'_{\mu\nu}=\td{g}_{\mu\nu}+\lie_\xi
\td{g}_{\mu\nu}$ does represent a new configuration if the
coordinate system is held fixed. For the corresponding physical
meaning of $\xi$, note a real physical coordinate system is
nothing but a lattice of observers (clocks and rulers). The
observed fluctuation in the metric can also be interpreted as
that, the metric is held fixed, but it is the lattice of observers
oscillate involuntarily, driven by quantum fluctuations of the
spacetime. In order for the observers to see a metric fluctuation
$\delta \td{g}_{\mu\nu} =\lie_\xi \td{g}_{\mu \nu}$, the
involuntary motion of the observers must be $\delta
x^\mu=-\xi^\mu$. As such, we immediately have the following
constraints on $\xi$:
\begin{itemize}
\item First of all, since $\xi^\mu$ corresponds to the involuntary
motion of observers, it should be finite;

\item Secondly, $\xi^r(r_0)\neq0$ means that the horizon either
expands or shrinks, and this in general leads to a different black
hole. So we must require $\xi^r(r_0)=0$;

\item Similarly, $\xi^i(r_0)\neq0$ in general changes the metric
on the horizon, and which should also be forbidden if we want to
talk about the same black hole.
\end{itemize}
With these considerations, one can expand $\xi^\mu$ near the black
hole horizon as
\be\xi^\mu=\sum_{k=0}^\infty\xi_{(k)}^\mu (r-r_0)^k\,,\quad
\xi_{(0)}^r=\xi_{(0)}^i=0\,, \label{anstaz.xi}\ee
where all the functions $\xi_{(k)}^\mu$ depend only on
$\theta^i,\phi^a$ and $t$.

Given this expansion, there are some further constraints that need
to be satisfied,
\begin{itemize}
\item Firstly, we need to determine under what conditions can a
perturbed configuration be regarded as the {\it same} to the
original black hole. The most straightforward choice seems to be
that the induced metric on the horizon should remain fixed. But
since the time $t$ is in many sense on the same footing as the
azimuthal angles $\phi^a$, we impose a stronger condition that the
induced metric on the $r=r_0$ hypersurface should remain fixed,
which means\footnote{Throughout the paper, we use ``$\approx$" to
relate quantities that are equal in the limit $r\rightarrow r_0$.}
\be\delta\td{g}_{ij}\approx \delta\td{g}_{iA} \approx \delta
\td{g}_{AB} \approx0\,,\quad \forall~~ i,j,A,B\,. \label{bd.r}\ee

\item Similarly, we also require that the volume density of the
full spacetime remains the same on the horizon,\footnote{Steve
Carlip told me that he had known the significance of this
condition for sometime, but he had been reluctant to use it for
lack of a good justification. Here we choose to use this condition
because (1) it is intuitively consistent with the requirement that
the black hole should remain the same on the horizon and (2) it is
technically very helpful.}
\be\delta\sqrt{-\td{g}}\approx0\,.\label{bd.g}\ee

\item Lastly, we require that there is no mixing between
$\theta^i$ and any other directions,
\be\delta\td{g}_{ir} \approx\delta\td{g}^{ir}
\approx\delta\td{g}^{ij} \approx\delta\td{g}^{iA} \approx0\,,\quad
\forall\;\,i,A\,.\label{bd.i}\ee
This is motivated by the desire to preserve $\theta^i$ as
longitudinal angles. (Otherwise one cannot say for sure that $r$
is the radial direction and $r=r_0$ is the horizon.) But beyond
that, there is not a very good reason for why one must do this. So
we will treat (\ref{bd.i}) as a hand-put-in condition.
\end{itemize}
To study the consequences of the boundary conditions (\ref{bd.r}),
(\ref{bd.g}) and (\ref{bd.i}), let's use (\ref{anstaz.xi}) and
write down the corresponding perturbation over the background
(\ref{metric.general}) explicitly. The results are
\bea\lie_\xi\td{g}_{rr}&\approx&\pd_r\Big(\frac{f}{\Delta'}
\xi_{(1)}^r \Big) +\frac{\xi_{(1)}^i\pd_if}{\Delta'}
+\frac{f}\Delta\xi_{(1)}^r\,,\label{v.rr}\\
\lie_\xi\td{g}_{rA}&\approx&\frac{f}{\Delta'}\pd_A
\xi_{(1)}^r+\td{g}_{AB}\xi_{(1)}^B\,,\label{v.rA}\\
\lie_\xi\td{g}_{ri}&\approx&\frac{f}{\Delta'}\pd_i \xi_{(1)}^r
+q_{ij}\xi_{(1)}^j\,,\quad
\lie_\xi\td{g}^{ri}\approx0\,,\label{v.ri}\\
\lie_\xi\td{g}_{ij}&\approx&\lie_\xi\td{g}^{ij}\approx0\,,\label{v.ij}\\
\lie_\xi\td{g}_{ia}&\approx&g_{ab}D_i\xi_{(0)}^b\,,\label{v.ia}\\
\lie_\xi\td{g}_{it}&\approx&-w_aD_i\xi_{(0)}^a\,,\label{v.it}\\
\lie_\xi\td{g}^{iA}&\approx&\frac{v^2}{f\Delta'}w^Aw^B\pd_B
\xi_{(1)}^i -q^{ij}\pd_j\xi_{(0)}^A\,,\label{v.iA}\\
\lie_\xi\td{g}_{ab}&\approx&g_{ac}D_b\xi_{(0)}^c
+g_{bc}D_a\xi_{(0)}^c\,,\label{v.ab}\\
\lie_\xi\td{g}_{at}&\approx&g_{ab}D_t\xi_{(0)}^b
-w_bD_a\xi_{(0)}^b\,,\label{v.at}\\
\lie_\xi\td{g}_{tt}&\approx&-2w_aD_t\xi_{(0)}^a\,,\label{v.tt}\\
\lie_\xi\sqrt{-\td{g}}&\approx& \sqrt{-\td{g}}\;\Big(\xi_{(1)}^r
+\pd_A\xi_{(0)}^A\Big)\,.\label{v.detg}\eea
where $D_\mu\xi_{(0)}^a\equiv\pd_\mu\xi_{(0)}^a -w^a \pd_\mu
\xi_{(0)}^t$. Comparing these results with (\ref{bd.r}),
(\ref{bd.g}) and (\ref{bd.i}), we find that
\bea D_i\xi_{(0)}^a\approx D_a\xi_{(0)}^a\approx D_t \xi_{(0)}^a
\approx0\,,\quad \xi_{(1)}^r=-\pd_A\xi_{(0)}^A\,,\nn\\
\xi_{(1)}^i=-q^{ij}\frac{f}{\Delta'}\pd_j\xi_{(1)}^r\,,\quad \pd_i
\xi_{(0)}^A= q_{ij}\frac{v^2}{f\Delta'}w^Aw^B\pd_B
\xi_{(1)}^j\,.\label{bd.equations}\eea
To solve these equations, note $D_\mu\xi_{(0)}^a\approx0$ are
easily solved with
\be\xi_{(0)}^a=\Omega^a \xi_{(0)}^t\,,\quad\Longrightarrow \quad
D_\mu\xi_{(0)}^a=(\Omega^a-w^a)\pd_\mu\xi_{(0)}^t \approx0\,,
\label{bd.so1}\ee
where $\Omega^a$ is defined below (\ref{wa.infty}). The other
equations are then uniquely solved by
\be\xi_{(1)}^r=-\pd_A\xi_{(0)}^A=-\Omega^A\pd_A\xi_{(0)}^t\,,
\quad \pd_i\xi_{(0)}^t=0\,,\quad \xi_{(1)}^i=0\,.\label{bd.so2}\ee
With these results, the only non-vanishing variation of the metric
elements are
\be\lie_\xi\td{g}_{rr}\approx\cO(\frac1\Delta)\,,\quad
\lie_\xi\td{g}_{rA}\approx\cO(\frac1{\Delta'})\,.\ee
Note the variation of $\td{g}_{rr}$ is of the same order as
$\td{g}_{rr}$ itself. Something similar also exists in the usual
Kerr/CFT correspondence \cite{ghss08}.

Now at the leading order $\xi^\mu$ depends only on $\xi_{(0)}^t$,
which is a free function of $\phi^a$ and $t$. One can expand
$\xi_{(0)}^t$ using the Fourier modes $e^{-im (\phi^{\bar a}
-\td\Omega^{\bar a} t)}$, where $\phi^{\bar a}$ is one of the
azimuthal angles, $m$ is an integer, and $\td\Omega^{\bar a}$ is a
constant to be determined. When appropriately normalized, we find
($\rho\equiv r-r_0$)
\be\bar{a}_m\equiv\xi^\mu\pd_\mu=-e^{-im (\phi^{\bar a} -\td
\Omega^{\bar a} t)}\Big\{ \Big[i \,m\,\rho +\cO(\rho^2) \Big]
\,\pd_r +\cO(\rho^2) \pd_i +\Big[\frac{\Omega^A}{\Omega^{\bar
a}-\td\Omega^{\bar a}}+\cO(\rho) \Big]\pd_A \Big\} \,.
\label{def.bar.am}\ee
As a comparison, the generators found in
\cite{carlip11,chen.zhang11} are
\be\bar{a}_m=-e^{-im (\phi^{\bar a}-\Omega^{\bar a}t)}\Big[i\,m
\,\rho\,\pd_r + \pd_{\bar a} +{\rm subleading~terms}\Big]\,.\ee
At the leading order, the generators (\ref{def.bar.am}) satisfy
the (centerless) Virasoro algebra,
\be i[\bar{a}_m\,,\,\bar{a}_n]=(m-n)\bar{a}_{m+n}\,.
\label{virasoro}\ee
Like in (\ref{def.E}) and (\ref{def.Ja}), one can define the
charge corresponding to $\bar{a}_m$ by using (\ref{app.def.Hxi2}),
\be\delta L_m^{\bar a} =-\int_{horizon}\Big(\delta
\bQ_{(\bar{a}_m)} -i_{(\bar{a}_m)} {\bf\Theta}_{\delta}
\Big)\,.\label{def.Lm}\ee
For $m\neq0$, we find because of the factor $e^{-im \phi^{ \bar
a}}$,
\be\bar\delta L_m^{\bar a}=0\,,\quad m=\pm1,\pm2,\cdots\,.\ee
So $L_m^{\bar a}\,,\;\forall\;m\neq0$ are independent on the
background metric (\ref{metric.general}). For $m=0$,
\be \bar{a}_0=-\frac{\breve\xi}{\Omega^{\bar a} -\td\Omega^{\bar
a}}+\cO(\rho)\,,\quad \breve\xi\equiv\Omega^A\pd_A=\pd_t+\Omega^a
\pd_a\,.\ee
In the case when $\bar\delta(\Omega^{\bar a}-\td\Omega^{\bar
a})=0$, we find using (\ref{def.E}) and (\ref{def.Ja})
\bea\bar\delta L_0^{\bar a}=-\int_{horizon}\bar\delta
\bQ_{(\bar{a}_0)} -i_{(\bar{a}_0)}{\bf\Theta}_\delta
=\bar\delta\Big(\frac{E-\Omega^aJ_a}{\Omega^{\bar a}
-\td\Omega^{\bar a}}\Big)\,,\nn\\
\Longrightarrow\quad L_0^{\bar a} =\frac{E-\Omega^a
J_a}{\Omega^{\bar a} -\td\Omega^{\bar a}}+({\rm
background~independent~constant})\,.\label{def.L0}\eea

\section{The non-extremal case}\label{sec.non.extremal}

Given the charges (\ref{def.Lm}), there is a central extension to
(\ref{virasoro}), just as described in (\ref{app.algebra}) and
(\ref{app.final.algebra}) in the appendix. The central charge can
be found through (\ref{app.virasoro}) and
(\ref{app.central.charge}). To find the result explicitly, let's
firstly look at the quantity defined in (\ref{app.central.term})
and (\ref{app.def.Kab}). The relevant component is (note
$\lie_m\equiv\lie_{\bar{a}_m}$)
\bea K^{tr}(\lie_n,\lie_m)&=&\frac1{16\pi}\Big\{-\frac{\td{h}}2
\bar{a}_m^{t r} +\td{h}^{t\rho} \td\nabla_\rho\bar{a}_m^r
-\td{h}^{r \rho} \td\nabla_\rho\bar{a}_m^t-(\td\nabla^t \td{h}^{r
\rho}-\td\nabla^r\td{h}^{t\rho})\bar{a}_{m\rho}\nn\\
&&\qquad+\bar{a}_m^t(\td\nabla_\rho \td{h}^{r\rho}-\td\nabla^r
\td{h}) -\bar{a}_m^r(\td\nabla_\rho \td{h}^{t\rho}-\td\nabla^t
\td{h})\Big\}\,.\eea
where $\td{h}_{ \mu \nu}\equiv\lie_n\td{g}_{\mu\nu}$. For later
convenience, we shall calculate the result for the more general
Virasoro generators,
\be\bar{a}_m=-e^{-im (\phi^{\bar a}-\hat\Omega^{\bar a} t)} \Big\{
\Big[i\,m\,\rho +\cO(\rho^2)\Big]\,\pd_r +\cO(\rho^2) \pd_i
+\Big[\chi^A +\cO(\rho) \Big]\pd_A\Big\}\,, \label{general.am}\ee
where $\hat\Omega^{\bar a}$ and $\chi^A$ are arbitrary constants,
except for
\be\chi^{\bar a}=1+\hat\Omega^{\bar a}\chi^t\,.
\label{chi.bar.a}\ee
Note in (\ref{def.bar.am}), we have
\be\hat\Omega^{\bar a}=\td\Omega^{\bar a}\,,\quad \chi^{\bar
a}=\frac{\Omega^{\bar a}}{\Omega^{\bar a}-\td\Omega^{\bar
a}}\,,\quad \chi^t=\frac1{\Omega^{\bar a}-\td\Omega^{\bar
a}}\,,\label{prop.bar.am}\ee
which obviously satisfy (\ref{chi.bar.a}).

Using (\ref{chi.bar.a}) and assuming $T\propto\Delta'(r_0)\neq0$,
we find
\be K^{tr}(\lie_{-m},\lie_m)\approx-\frac{4im^3}{16 \pi} \Big[
\chi^t(\Omega^{\bar a}-\hat\Omega^{\bar a})-\frac12\Big]\frac{v^2
(\Omega^{\bar a}-\hat\Omega^{\bar a})}{f\Delta'}+\cdots\,,
\label{central.1}\ee
where we have only kept terms that are third order polynomials in
$m$. The omitted terms are all finite and are linear in $m$. Note
again that ``$\approx$" means equal on the horizon. The subleading
terms (those of $\cO(\rho^2)$ for $\pd_r$ and $\pd_i$, and those
of $\cO(\rho)$ for $\pd_A$) are not constrained in
(\ref{def.bar.am}). So it is reassuring to note that the
corresponding subleading terms in (\ref{general.am}) have {\it no}
contribution to (\ref{central.1}) either.

Now using (\ref{app.central.charge}), the central charge is
\bea c^{\bar a}&=&\frac{3}\pi\int_{horizon} (d^{D-2}x)_{tr}2
\sqrt{-\td{g}}\;\Big[\chi^t (\Omega^{\bar a}-\hat\Omega^{\bar a})
-\frac12\Big] \frac{v^2(\Omega^{\bar a}-\hat\Omega^{\bar a})}{f\Delta'}\nn\\
&=&\frac{3}\pi\int_{horizon} (d^{D-2}x)_{tr}2\sqrt{q g}\;\Big[
\chi^t(\Omega^{\bar a}-\hat\Omega^{\bar a})-\frac12\Big]\frac{v_0
(r_0)(\Omega^{\bar a}-\hat\Omega^{\bar a})}{\Delta'(r_0)}\nn\\
&=&\frac{3}{\pi^2}\Big[\chi^t(\Omega^{\bar a}-\hat\Omega^{\bar a})
-\frac12 \Big]\cdot\frac{\Omega^{\bar a}-\hat\Omega^{\bar a}}{T}
\cdot S\;,\label{central.c1} \eea
where $T$ is the temperature (\ref{temperature}) and $S$ is the
entropy (\ref{first.law}). For the generators (\ref{def.bar.am}),
we can use (\ref{prop.bar.am}) to further reduce the result to
\be c^{\bar a}=\frac{3}{2\pi^2}\cdot\frac{\Omega^{\bar a}
-\td\Omega^{\bar a}}{T} \cdot S\,.\ee
Note in order for $c^{\bar a}$ to be non-negative, we need
$\Omega^{\bar a}\geq\td\Omega^{\bar a}$. On the other hand, the
Frolov-Thorne temperature for the modes $e^{-im(\phi^{\bar
a}-\td\Omega^{\bar a}t)}$ is \cite{ghss08,carlip11}
\be T^{\bar a}=\frac{T}{\td\Omega^{\bar a}-\Omega^{\bar a}}\,,
\label{def.Ta}\ee
which is negative for $\Omega^{\bar a}>\td\Omega^{\bar a}$. If we
want both $c^{\bar a}$ and $T^{\bar a}$ to be non-negative, then
the only choice is $\td\Omega^{\bar a} =\Omega^{\bar
a}$.\footnote{The different signs between $c^{\bar a}$ and
$T^{\bar a}$ at $\td\Omega^{\bar a}\neq \Omega^{\bar a}$ can be
understood from the fact that the horizon of
(\ref{metric.general}) is a {\it frozen} surface. That is, for an
observer comoving with the horizon, there should not be any
propagating signals in the $\theta^i$ or $\phi^a$ directions on
the horizon (because such signals are always space-like, i.e.,
superluminal). So on the horizon, all perturbations can only be a
function of $\phi^a -\Omega^a t$. As a consequence, the modes
$e^{-im(\phi^{\bar a}-\Omega^{\bar a}t)}$ should in fact be
understood as $e^{-im(\phi^{\bar a}-\Omega^{\bar a}t)} =e^{-im
(\phi^{\bar a}-\Omega^{\bar a}t)}\prod_{a\neq\bar{a}} e^{-i0
(\phi^a -\Omega^a t)}$.} In this case, $c^{\bar a}$ vanishes and
$T^{\bar a}$ diverges. The generators (\ref{def.bar.am}) also
diverge. As suggested in \cite{carlip11}, the origin of such
singular behaviors could be physical. Using the canonical version
of Cardy's formula, we find
\be S^{\bar a}=\frac{\pi^2}3c^{\bar a} T^{\bar a}=\frac{S}2\,.
\label{entropy.c1}\ee
This result resembles that in section 2 of \cite{carlip11} in a
remarkable way. Note Cardy's formula only gives half the
Bekenstein-Hawking entropy. We will discuss more about this issue
when we conclude.

Note the above result does not depend on which azimuthal angle is
used. This is similar to what happens in the case of extremal
Kerr/CFT correspondence \cite{chow.cvetic.lu.pope09}. Although we
have the same number of Virasoro algebras as that of the azimuthal
angles, they are not independent in terms of counting the degrees
of freedom for the black hole.

\section{The extremal case}\label{sec.extremal}

As mentioned in the previous section, the subleading terms in
(\ref{general.am}) have no contribution to the central term
(\ref{central.1}). Unfortunately, this is not true for extremal
black holes. In order to talk sensibly about the central charges
for extremal black holes, one must further require that
(\ref{general.am}) obey the Virasoro algebra up to the sub-leading
order. We then find
\bea\bar{a}_m&=&-e^{-im (\phi^{\bar a}-\hat\Omega^{\bar a} t)}
\Bigg(\Big\{i\,m\,\rho+\Big[mu^r+\frac{im^2}2(u^{\bar a}
-\Omega^{\bar a}u^t)\Big]\rho^2 +\cO(\rho^3)\Big\}\pd_r\nn\\
&&\qquad\qquad\qquad\quad+\Big[mu^i\rho^2+\cO(\rho^3)\Big]
\pd_i+\Big[\chi^A+mu^A\rho+\cO(\rho^2)\Big]\pd_A\Bigg)\,,
\label{general.am2}\eea
where $u^r$, $u^i$ and $u^A$ are free functions of $\theta^i$.
With (\ref{general.am2}), we find that the contribution of the
subleading terms to the central charge vanishes again. The
following result is valid for both extremal and non-extremal black
holes,
\be
K^{tr}(\lie_{-m},\lie_m)\approx-m^3\Big(\frac{\Delta'}{\Delta^2}
Z_1 +\frac{Z_2}\Delta\Big)+\cdots\,,\label{central.general}\ee
where $``\cdots"$ denotes terms linear in $m$ and
\bea Z_1&=&-\frac{2iv_0^2(r_0)(\Omega^{\bar a}-\hat\Omega^{\bar
a}) }{f(r_0,\theta^i)}\rho^2+\Big[-\frac{2iv_0^2(r_0)w'^{\bar a}
(r_0)}{f(r_0,\theta^i)} +(\Omega^{\bar a}-\hat\Omega^{\bar
a})G_1(r_0,\theta^i)\Big]\rho^3+\cO(\rho^4)\,,\nn\\
Z_2&=&\frac{4iv_0^2(r_0)(\Omega^{\bar a}-\hat\Omega^{\bar a})^2
\chi^t}{f(r_0,\theta^i)}\rho+\Big[\frac{2iv_0^2(r_0)w'^{\bar a}
(r_0)}{f(r_0,\theta^i)} +(\Omega^{\bar a}-\hat\Omega^{\bar
a})G_2(r_0,\theta^i)\Big]\rho^2+\cO(\rho^3)\,.
\label{central.general.functions}\eea
The detail of the two functions $G_1$ and $G_2$ will not be
important for us. We only need to know that they are both of order
$\cO(1)$. Note we have preserved the dependence of
(\ref{central.general}) and (\ref{central.general.functions}) on
$r$ only through $\Delta$, $\Delta'$ and $\rho(=r-r_0)$.

One can check that (\ref{central.general}) reduces to
(\ref{central.1}) in the non-extremal case, $\Delta'(r_0)\neq0$.
In the extremal case $\Delta'(r_0)=0$,
\bea&&K^{tr}(\lie_{-m},\lie_m)\approx\frac{4i m^3 v_0^2(r_0)
w_0'^{\bar a}(r_0)}{16\pi\Delta''(r_0)f(r_0,\theta^i)}(1+ G)+\cdots\,,\nn\\
&&G=\frac{\Omega^{\bar a}-\hat\Omega^{\bar a}}{w_0'^{\bar a}(r_0)}
\Big\{\frac2\rho\Big[1-\chi^t(\Omega^{\bar a}-\hat\Omega^{\bar
a})\Big] +\frac{2\Delta'''(r_0) }{3\Delta''(r_0)} \Big[\chi^t
(\Omega^{\bar a} -\hat\Omega^{\bar a})-\frac12\Big]\nn\\
&&\qquad\qquad\qquad~~-\frac{f(r_0,\theta^i)}{2iv_0^2
(r_0)}(2G_1+G_2)\Big\}\,, \label{central.2}\eea
The first term in $G$ diverges as $\rho\rightarrow0$. But this
divergence will go away once we use (\ref{prop.bar.am}). In fact,
as was discussed in the previous section, $\td\Omega^{\bar
a}=\Omega^{\bar a}$. So $G=0$ upon using (\ref{prop.bar.am}). From
(\ref{app.central.charge}), the central charge for the extremal
case is
\bea c^{\bar a}&=&-\frac3\pi\int_{horizon}(d^{D-2}x)_{tr}2
\sqrt{-\td{g}}\;\frac{v_0^2(r_0) w_0'^{\bar a}(r_0)}{\Delta''
(r_0) f(r_0,\theta^i)}=\frac3{\pi^2}\frac{S}{\wtd{T}^{\bar
a}}\,,\label{central.c2}\eea
where $\wtd{T}^{\bar a}$ is the Frolov-Thorne temperature related
to $\phi^{\bar a}$ from the usual Kerr/CFT calculation
\cite{mei10},
\be \wtd{T}^{\bar a}=-\frac{\Delta''(r_0)}{4\pi v_0(r_0) w'^{\bar
a}(r_0)}\,.\ee
In the present context, (\ref{def.Ta}) is valid for both the
extremal and non-extremal cases. But $T^{\bar a}$ is now
indefinite because $T=\Omega^{\bar a}-\td \Omega^{\bar a}=0$. As
suggested in \cite{carlip11}, one can identify $T^{\bar a}$ with
$\wtd{T}^{\bar a}$. In this case, Cardy's formula gives
\be S^{\bar a}=\frac{\pi^2}3 c^{\bar a} T^{\bar a}=S\,,
\label{entropy.c2}\ee
which is exactly the Bekenstein-Hawking entropy.

As a side remark, note the contribution to the central charge
(\ref{central.general}) comes from different terms for extremal
and non-extremal cases. So in terms of the central charge, one
cannot expect a smooth transition from the non-extremal case to
the extremal case. This is another indication of the discontinuity
that arises in taking the extremal limit for non-extremal black
holes (see, e.g. \cite{carroll.johnson.randall09}).

\section{Summary}\label{sec.summary}

To summarize, we have studied conformal symmetries one the horizon
of a general stationary and axisymmetric black hole. We find that
consistent and physically reasonable boundary conditions exist,
which uniquely determine a set of symmetry generators that form a
copy of the Virasoro algebra. The construction is designed for
black holes in arbitrary spacetime dimensions and in arbitrary
theories. But for practical reasons, explicit calculation is only
done for Einstein gravity plus a (possibly zero) cosmological
constant. We find that one can deduce the full Bekenstein-Hawking
entropy by using Cardy's formula for extremal black holes. For
non-extremal black holes, Cardy's formula only gives half the
Bekenstein-Hawking entropy.

As a possible explanation to the failure in the non-extremal case,
we note that our boundary conditions (\ref{bd.r}), (\ref{bd.g})
and (\ref{bd.i}) are very stringent, and it uniquely allows for
only one copy of the Virasoro algebra. In contrast, such as
indicated in \cite{cms10}, the dual CFT for a non-extremal black
hole could be non-chiral and there might be two copies of Virasoro
algebras. This means that there might be a second Virasoro algebra
but which is filtered out by (\ref{bd.r}), (\ref{bd.g}) and
(\ref{bd.i}). However, so far we have not been able to find the
more general boundary conditions that allow for two copies of the
Virasoro algebras.

On the other hand, it is also possible that the problem is due to
something else. One indication is that, as mentioned before, the
horizon is a {\it frozen} surface, and all ``fluctuations" on the
horizon can only be a function of $\phi^a-\Omega^at$. From this
perspective, it seems very unlikely that there could be a second
independent copy of the Virasoro algebra, because we do not have a
second independent coordinate to work with.\footnote{This is to be
contrasted with the asymptotic symmetries at spatial infinity of a
BTZ black holes, where it is natural to define two independent
coordinates like $\phi\pm t/\ell$ ($\ell$ is the $AdS$ radius),
and which are both periodic in $\phi$.} So it is possible that
(\ref{def.bar.am}) is all we have on a black hole horizon. For
extremal black holes, it is already with some luck that Cardy's
formula does reproduce the full Bekenstein-Hawking entropy. And in
fact no one knows why this must work. For non-extremal black
holes, it is then conceivable that one may have to go beyond
simply applying Cardy's formula to deduce the full black hole
entropy.

Another issue with non-extremal black holes is that the central
charge and the Frolov-Thorne temperature are singular. However,
here one is not sure if the singular behavior is intrinsic to the
problem, or if one can find a better alternative construction that
does not have such singular behaviors. We wish to understand this
issue better in the future.

Despite such problems with the non-extremal case, we note Cardy's
formula does give qualitatively correct result for the entropy.
More importantly, it is remarkable that although our boundary
conditions (\ref{bd.r}), (\ref{bd.g}) and (\ref{bd.i}) are as
stringent as one can imagine, they still allow for non-trivial
physical results. What's more, our boundary conditions are imposed
on the metric elements on the horizon directly, and we do not need
to introduce an intermediate stretched horizon. So if the boundary
conditions (\ref{bd.r}), (\ref{bd.g}) and (\ref{bd.i}) are truly
physically relevant, then quantum fluctuations near the horizon
must be generated by (\ref{def.bar.am}), making it more convincing
that quantum gravity on a black hole horizon is dual to some 2D
conformal field theory.

\section*{Acknowledgement}

I thank Stefan Theisen for many helpful discussions and for
reading the manuscript. I also thank Steve Carlip for
correspondence and for his penetrating comments and suggestions on
the draft. And I thank Dumitru Astefanesei for a discussion. This
work was supported by the Alexander von Humboldt-Foundation.

\appendix

\section{The covariant phase space method}\label{sec.app}

In this section we compile all the necessary tools that are needed
in the bulk discussion of the paper. Although we do re-drive some
of the formulae that are particularly important for us, there is
nothing essentially new here. For original works on the covariant
phase space method, one can consult \cite{abbott.deser81,wald93,
iyer.wald94,anderson.torre96, torre97,barnich.brandt01,
barnich.compere07}. For earlier works on using the method to
calculate central charges for black holes, one can see
\cite{carlip99,koga01,silva02}. Part of our description follow
\cite{wald93,iyer.wald94,wald09} closely.

As a motivating example, we start with the case of one-dimensional
motion in classical mechanics. The lagrangian is $L=L(q, \dtq)$,
with $q=q(t)$. Under a general operation $\hat\delta$ on
$q$,\footnote{Note $\hat\delta$ can either be a usual variation of
$q$, or of some other types such as a time derivative on $q$.}
\bea\hat\delta L&=&\Big(\frac{\pd L}{\pd q} -\frac{d}{dt}\frac{\pd
L}{\pd\dtq}\Big)\hat\delta q +\frac{d}{dt}\Big(\frac{\pd L}{\pd
\dtq}\hat\delta q\Big)=E\hat\delta q+\frac{d}{dt}\Theta_{\hat\delta}\,,\nn\\
E&=&\frac{\pd L}{\pd q} -\dtp\,,\quad p=\frac{\pd L}{\pd\dtq} \,,
\quad \Theta_{\hat\delta}=p\hat\delta q\,.\eea
In the canonical phase space, $p$ and $q$ are treated as
independent ``coordinates", and one may denote them as $z^1=q$ and
$z^2=p$. The Poisson bracket of any two functions $f=f(q,p)$ and
$g=g(q,p)$ can be written as
\bea\Big\{f\,,\,g\Big\}_{P.B.}=\frac{\pd f}{\pd q} \frac{\pd
g}{\pd p} -\frac{\pd f}{\pd p}\frac{\pd g}{\pd q}=\Omega^{mn}
\frac{\pd f}{\pd z^m}\frac{\pd g}{\pd z^n}\,,\quad m,n=1,2\,,\nn\\
(\Omega^{mn})=\left(\begin{matrix} &1\cr -1&\end{matrix}\right)\,,
\quad\Longrightarrow\quad (\Omega_{mn})=\left(\begin{matrix}
&-1\cr1&\end{matrix} \right)\,.\eea
For two arbitrary operations (say, $\hat\delta_1$ and
$\hat\delta_2$), one can also define the presymplectic potential
as
\be\Omega(\hat\delta_1,\hat\delta_2)\equiv\hat\delta_1
\Theta_{\hat\delta_2} -\hat\delta_2 \Theta_{\hat\delta_1}
=\hat\delta_1p\hat\delta_2q-\hat\delta_2p \hat\delta_1q
=\Omega_{mn}\hat\delta_1 z^m \hat\delta_2 z^n \,.\ee
An interesting way to define the Hamiltonian is to take
$\hat\delta_1$ to be a particular variation $\delta$, which takes
one solution to a nearby one, and to take
$\hat\delta_2=\frac{d}{dt}$,
\be\delta H=\Omega\Big(\delta,\frac{d}{dt}\Big)=\delta\Theta_{
(\frac{d}{dt})} -\frac{d}{dt}\Theta_\delta=\delta p\dtq -\dtp
\delta q\,.\ee
The Hamilton-Jacobi equations then follow in a straightforward
manor,
\be\dtq=\frac{\delta H}{\delta p}\,,\quad \dtp=-\frac{\delta
H}{\delta q}\,.\ee

In parallel, one can do the same for a general system. Denoting
the generalized coordinates of the canonical phase space as $z^m$,
$m=1,2,\cdots$, one defines the presymplectic potential as
\be \Omega(\hat\delta_1,\hat\delta_2) =\Omega_{mn}\hat\delta_1
z^m\hat\delta_2z^n\,.\ee
For any symmetric transformations $\hat\delta_\xi$, the variation
of the corresponding charge $H_\xi$ is
\be\delta H_\xi=\Omega(\delta,\hat\delta_\xi) =\Omega_{mn} \delta
z^m\hat\delta_\xi z^n\,,\label{app.def.Hxi1}\ee
where $\delta$ is again the particular variation that takes one
solution to a nearby one. The Poisson bracket (or more generally
the Dirac bracket) of any two such charges is
\be\Big\{H_\xi\,,\,H_\zeta\Big\}=\Omega^{mn}\frac{\delta
H_\xi}{\delta z^m} \frac{\delta H_\zeta}{\delta z^n}=
-\Omega(\hat\delta_\xi, \hat\delta_\zeta)=\Omega(\hat
\delta_\zeta, \hat\delta_\xi)\,.\label{app.def.bracket}\ee
Using the Jacobi identity and an arbitrary charge $Q$, one can
further derive that
\bea\{Q,H_{[\xi,\zeta]}\}&=&\hat\delta_{[\xi, \zeta]}Q =(\hat
\delta_\xi\hat\delta_\zeta-\hat\delta_\zeta\hat\delta_\xi)Q
=\hat\delta_\xi\{Q,H_\zeta\}-\hat\delta_\zeta\{Q,H_\xi\}\nn\\
&=&\{\{Q,H_\zeta\},H_\xi\} -\{\{Q,H_\xi\},H_\zeta\}
=\{\{H_\xi,H_\zeta\},Q\}\,,\nn\\
\Longrightarrow\quad\{H_\xi,H_\zeta\}&=&-H_{[\xi,\zeta]}
+K_{[\xi,\zeta]}\,,\quad{\rm where}\quad \{K_{[\xi,\zeta]}
\,,\,Q\} =0\,,\quad \forall~Q\,.\label{app.algebra}\eea
Hence, simply because of the Jacobi identity, the algebra of the
Poisson (Dirac) bracket is isomorphic (up to a possible central
extension $K_{[\xi,\zeta]}$) to the Lie algebra $\hat\delta_\xi
\hat\delta_\zeta-\hat\delta_\zeta\hat\delta_\xi=\hat\delta_{[\xi,
\zeta]}$.

Now consider the general action,
\be S=\int_\cM\bL\,,\quad \bL=\cL(\Phi, \pd_\mu\Phi, \pd_\mu
\pd_\nu\Phi,\cdots)\ast\one\,,\ee
where $\Phi$ denotes all possible fields collectively. From now
on, a bold faced letter such as $\bL$ always stands for a
differential form. Under an arbitrary operation $\hat\delta$ on
the fields,
\be\hat\delta\bL=(\hat\delta\Phi)E_\Phi\ast\one +d{\bf
\Theta}_{\hat\delta}\,,\ee
where all terms involving a derivative on $\hat\delta\Phi$ have
been moved into $d\bf\Theta_{\hat\delta}$. The Euler-Lagrange
equations are just $E_\Phi=0$. In the special case when
$\hat\delta$ is identified with a Lie derivative $\lie_\xi=d\cdot
i_\xi +i_\xi\cdot d$,
\bea\lie_\xi\bL=d(i_\xi\bL)=(\lie_\xi\Phi)E_\Phi\ast\one +d{\bf
\Theta}_\xi\,,\quad\bJ_\xi ={\bf\Theta}_\xi -i_\xi\bL\,,\nn\\
\Longrightarrow\quad d\bJ_\xi=-(\lie_\xi\Phi)E_\Phi\ast\one
\cong0\,,\quad\Longrightarrow\quad \bJ_\xi\cong d\bQ_\xi\,,
\label{def.Jxi}\eea
where $``\cong"$ means equal after using the equations of motion
$E_\Phi=0$. For a Killing vector $\xi$, one may call $\bJ_\xi$ the
corresponding Noether current. Now consider the variation
$\hat\delta=\delta$ that takes a classical solution to a nearby
one,
\be\delta\bJ_\xi=\delta{\bf\Theta}_\xi -\delta (i_\xi\bL)
=\delta{\bf\Theta}_\xi-i_\xi\cdot d{\bf \Theta}_{\delta}
=\bw(\delta, \lie_\xi) +d(i_\xi{\bf\Theta}_{\delta})\,,
\label{def.bw}\ee
where $\bw(\delta,\lie_\xi) \equiv\delta \bf\Theta_\xi -\lie_\xi
\bf\Theta_\delta$. Since $\delta$ takes one solution to a nearby
one, $\bJ_\xi$ stays exact as in (\ref{def.Jxi}). As a result,
\be\delta\bJ_\xi=d\delta\bQ_\xi\,,\quad\Longrightarrow \quad
\bw(\delta,\lie_\xi)=d\bk(\delta,\lie_\xi)\,, \quad
\bk(\delta,\lie_\xi)\equiv\delta\bQ_\xi -i_\xi
{\bf\Theta}_{\delta}\,.\ee
In the case when $\xi$ is a Killing vector,
\be\lie_\xi=0\quad \Longrightarrow\quad \bw(\delta,
\lie_\xi)=0\,,\quad \Longrightarrow\quad 0=\int_V\bw(
\delta,\lie_\xi)=\oint_{\pd V} \bk(\delta,\lie_\xi)
\,,\label{app.indentity1}\ee
where $V$ is a Cauchy surface. We are particularly interested in
the spacetime of a stationary black hole, where one can take $V$
to be the space outside the horizon. As a result, $\pd V$ has two
disconnect pieces: one at spatial infinity and one at the horizon,
\be\oint_{\pd V}=\int_{+\infty}-\int_{Horizon}\,.
\label{app.indentity2}\ee
Usually one defines the charge corresponding to $\lie_\xi$ through
an integral at spatial infinity,
\be\delta H_\xi=\int_{+\infty} \bk(\delta, \lie_\xi) =\int_{+
\infty} \delta\bQ_\xi -i_\xi {\bf\Theta}_{\delta}\,. \ee
But because of (\ref{app.indentity1}) and (\ref{app.indentity2}),
this is equivalent to defining
\be\delta H_\xi=\int_{horizon}\bk(\delta,\lie_\xi)
=\int_{horizon}\delta\bQ_\xi -i_\xi {\bf\Theta}_{\delta}\,.
\label{app.def.Hxi2}\ee
It is the second definition that we want to us in this paper. It
is also straightforward to generalize such a definition to charges
of boundary symmetries.

Comparing (\ref{app.def.Hxi2}) with (\ref{app.def.Hxi1}), we see
that in the present discussion, $\hat\delta_\xi=\lie_\xi$ and
\be\Omega(\delta,\lie_\xi)=\int_{horizon}\bk(\delta,\lie_\xi)
\,.\ee
But this result is not enough for us to recover the full
presymplectic potential $\Omega(\lie_{\xi_1},\lie_{\xi_2})$, which
is needed to define the Poisson/Dirac bracket
(\ref{app.def.bracket}) for the corresponding charges. One the
other had, one can still define the following quantity,
\be K[\lie_\zeta,\lie_\xi]\equiv\Omega(\delta,\lie_\xi)|_{\delta
\rightarrow\lie_\zeta}=\int_{horizon}\bk(\delta,\lie_\xi)|_{
\delta\rightarrow\lie_\zeta}\,. \label{app.central.term}\ee
And one can write
\bea\Big\{H_\xi\,,\,H_\zeta\Big\}=\Omega(\lie_\zeta,\lie_\xi)
=-H[\lie_\zeta,\lie_\xi] +K[\lie_\zeta,\lie_\xi]\,,
\label{app.final.algebra}\\
H[\lie_\zeta,\lie_\xi]=K[\lie_\zeta,\lie_\xi]-\Omega(\lie_\zeta,
\lie_\xi)\,.\eea
Although (\ref{app.final.algebra}) looks like (\ref{app.algebra})
very much, it is not guaranteed that
\be H_{[\xi,\zeta]}=H[\lie_\zeta,\lie_\xi]\,,\quad
~K_{[\xi,\zeta]}=K[\lie_\zeta,\lie_\xi]\,.\ee
But rigorous treatment (see, e.g. \cite{barnich.brandt01,
barnich.compere07}) does show that $K[\lie_\zeta,\lie_\xi]$
contains all the information about $K_{[\xi,\zeta]}$. This is good
enough for our purpose here. Now if the Virasoro algebra
\be\Big[L_m\,,\,L_n\Big]=(m-n)L_{m+n} +\frac{c}{12}m(m^2-1)
\delta_{m+n}\,,\label{app.virasoro}\ee
is realized from (\ref{app.final.algebra}) through the canonical
quantization $i\{\cdot,\cdot\} \rightarrow[\cdot, \cdot]$, then
one can read off the central charge from the coefficient of $m^3$
in the term $K[\lie_{-m},\lie_m]$,
\be c=12i\Big({\rm the~coefficient~of}~m^3{\rm~in} ~K[\lie_{-m},
\lie_m]\Big)\,.\label{app.central.charge}\ee

In the following, let's calculate $\bk(\delta,\lie_\xi)$ for
Einstein gravity plus a (possibly zero) cosmological constant. To
simplify notations, we will drop the ``tilde" from the full metric
in (\ref{elements.metric}) from now on. For differential forms, we
use the notation
\be(d^{D-p}x)_{\mu_1\cdots\mu_p}\equiv\frac1{p!(D-p)!}
\varepsilon_{\mu_1\cdots\mu_p\nu_1\cdots\nu_{D-p}^{~}}
dx^{\nu_1}\wedge\cdots\wedge dx^{\nu_{D-p}^{~}}\,,\quad
|\varepsilon_{\cdots}|=1\,,\ee
with which the Hodge-$\ast$ dual of a $p$-form $\bw_p=\frac1{p!}
w_{\mu_1 \cdots \mu_p}dx^{\mu_1}\wedge\cdots\wedge dx^{\mu_p}$ can
be written as
\be\ast\bw_p=\sqrt{|g|}\;(d^{D-p}x)_{\mu_1\cdots\mu_p}
w^{\mu_1\cdots\mu_p}\,,\quad\Longrightarrow\quad
\ast\one=\sqrt{|g|}\;d^Dx\,.\ee
For the exterior and interior products, one then has
\bea d\ast\bw_p=\sqrt{|g|}\;(d^{D-p+1}x)_{\mu_1\cdots
\mu_{p-1}}\nabla_{\mu_p}w^{\mu_1\cdots\mu_p}\,,\nn\\
i_\xi(d^{D-p}x)_{\mu_1\cdots\mu_p}=(d^{D-p-1}x)_{\mu_1
\cdots\mu_p\mu}(p+1)\xi^\mu\,.\eea
Now the action is
\be \bL=\Big(\frac{R -2\Lambda}{16\pi}\Big)\ast\one\,.
\label{app.action}\ee
Under an arbitrary operation $\hat\delta$ on the fields,
\bea\hat\delta\bL=\frac1{16\pi}\Big\{\frac{\hat h}2(R-2\Lambda)
+(-R^{\mu\nu}+\nabla^\mu\nabla^\nu-\nabla^2g^{\mu\nu})\hat{h}_{
\mu\nu}\Big\}\ast\one\,,\nn\\
\Longrightarrow\quad E^{\mu\nu}=\frac1{16\pi}\Big[\frac12
g^{\mu\nu}(R-2\Lambda)-R^{\mu\nu}\Big]=0\,,\nn\\
{\bf\Theta}_{\hat\delta}=\sqrt{-g}\;(d^{D-1}x)_\mu
\Big(\frac{\nabla_\nu\hat{h}^{\mu\nu} -\nabla^\mu
\hat{h}}{16\pi}\Big)\,,\label{app.Theta.delta}\eea
where $\hat{h}_{\mu\nu}\equiv\hat\delta g_{\mu\nu}$. In the case
when $\hat\delta=\lie_\xi$,
\bea\bJ_\xi&=&{\bf\Theta}_\xi-i_\xi\bL=\sqrt{-g}\;(d^{D-1} x)_\mu
\Big\{\frac{-\nabla_\nu\xi^{\mu\nu} +2R^{\mu\nu} \xi_\nu}{16\pi}
-\Big(\frac{R-2\Lambda}{16\pi}\Big)\xi^\mu\Big\}\nn\\
&=&\sqrt{-g}\;(d^{D-1}x)_\mu\Big(\frac{-\nabla_\nu
\xi^{\mu\nu}}{16\pi} \Big)=d\bQ_\xi\,,\nn\\
&\Longrightarrow&\bQ_\xi=\sqrt{-g}\; (d^{D-2}x)_{\mu\nu}
\Big(\frac{-\xi^{\mu\nu}}{16\pi}\Big)\,,\quad \xi^{\mu\nu}
=\nabla^\mu\xi^\nu -\nabla^\nu \xi^\mu\,. \label{app.def.Qxi}\eea
One can further derive that for $\bk(\delta,\lie_\xi)=
\delta\bQ_\xi-i_\xi{\bf\Theta}_{\delta}=\sqrt{-g}\;(d^{n-2}
x)_{\mu\nu}K^{\mu\nu}(\delta,\lie_\xi)$,
\bea K^{\mu\nu}(\delta,\lie_\xi)&=&\frac1{16\pi}\Big\{ -\frac{
\delta(\sqrt{-g}\;\xi^{\mu\nu})}{\sqrt{-g}}+\xi^\mu(\nabla_\rho
h^{\nu\rho}-\nabla^\nu h) -\xi^\nu(\nabla_\rho h^{\mu\rho}
-\nabla^\mu h)\Big\}\,,\nn\\
&=&\frac1{16\pi}\Big\{-\frac{h}2 \xi^{\mu \nu} +h^{\mu \rho}
\nabla_\rho\xi^\nu-h^{\nu\rho}\nabla_\rho\xi^\mu-(\nabla^\mu
h^{\nu \rho}-\nabla^\nu h^{\mu\rho})\xi_\rho\nn\\
&&\qquad+\xi^\mu(\nabla_\rho h^{\nu\rho}-\nabla^\nu h)
-\xi^\nu(\nabla_\rho h^{\mu\rho}-\nabla^\mu h)\Big\}\,.
\label{app.def.Kab}\eea

\end{document}